\documentclass[final,5p,times,twocolumn]{elsarticle}

\usepackage{amssymb}

\journal{Physics Letters B}

\begin{document}

\begin{frontmatter}



\title{Deformed one-quasiparticle states in covariant density 
functional theory}


\author{A.\ V.\ Afanasjev\corref{afansjev@erc.msstate.edu} and 
S.\ Shawaqfeh\fnref{Deceased}}

\address{Department of Physics and Astronomy, Mississippi State
University, MS 39762}

\begin{abstract}
  Systematic investigation of the accuracy of the description of the 
energies of deformed one-quasiparticle states has been performed in 
covariant density functional theory in actinide and rare-earth mass 
regions. The sources of the discrepancies between theory and experiment
are analyzed. Although some improvements in the description of ground 
state configurations and one-quasiparticle spectra can be achieved 
by better parametrization of the relativistic mean field Lagrangian, 
the analysis suggests that spectroscopic quality of their description 
can be achieved only in theoretical framework which takes into account 
particle-vibration coupling.
\end{abstract}

\begin{keyword}
One-quasiparticle states\sep covariant density functional theory\sep deformation
\sep particle-vibration coupling
\PACS 21.10.-k\sep 21.10.Pc\sep 21.60.Jz\sep 27.70.+q\sep 27.90.+b
\end{keyword}

\end{frontmatter}










\section{Introduction}

 Further progress in understanding low energy nuclear phenomena
in stable and exotic nuclei is strongly connected with the 
development of nuclear density functional theory (DFT) in its 
non-relativistic and relativistic (covariant) incarnations. 
These theories provide rather successful description of different 
aspects, such as deformations, masses, collective excitations 
etc, of nuclear many-body problem, see Refs.\ \cite{BHR.03,VRAL.05} 
and references therein. In addition, they are aimed on global 
(i.e. across full nuclear chart) description of the nuclear 
properties.

  However, absolute majority of the applications of nuclear DFT has 
been focused on collective properties of nuclei. This is due to the 
fact that time-odd mean fields are needed for the description of 
one-(many)-particle configurations which are characterized by broken
time-reversal symmetry in intrinsic frame (see Ref.\ \cite{VRAL.05}
and references therein). As a result, the description of such 
configurations is more complicated as compared with the one of the 
ground states of even-even nuclei.

  There are only few features of deformed nuclear systems, 
strongly dependent on single-particle degrees of freedom, which 
have been addressed in the DFT studies on the mean-field level 
and compared with experiment. When discussing the impact of
single-particle degrees of freedom on nuclear properties, it 
is important to separate different aspects of their physics 
which only weakly depend on each other. These are (i) deformation 
polarization effects induced by particle or hole, (ii) alignment 
properties of single-particle orbital in rotating potential and 
relevant polarization effects in time-odd mean fields, and 
(iii) the single-particle energies.

  Let us consider each of those aspects. The addition of the particle 
to (or the creation of the hole in) even-even core induces deformation 
polarization effects. The investigation of superdeformed rotational 
bands in the $A\sim 140-150$ mass region, which are one of the best 
examples of undisturbed single-particle motion \cite{SDDN.96,ALR.98}, 
has revealed that deformation  polarization effects induced by particle 
or hole are well described in nonrelativistic \cite{SDDN.96,Dy151} 
and covariant \cite{Dy151,ALR.98} DFT's; the average deviation from
available experimental differential quadrupole moments is around 20\%.
Similar (but somewhat less accurate because of the role of pairing) 
results have been obtained 
also in the $A\sim 130$ mass region of high- and superdeformation 
\cite{L.02,MADLN.07}. Macroscopic+microscopic (MM) method based on 
the Nilsson potential describes deformation polarization effects 
reasonably well but suffers from the fact that these effects are
not uniquely defined \cite{KR.98,Eet.00}.

  Alignment properties of single-particle orbital in rotating potential
can be accessed via effective (relative) alignments \cite{Rag.93} of two 
compared  rotational bands. This quantity sensitively depends on both the 
alignment properties of single-particle orbital by which two bands differ 
and polarization effects (mostly in time-odd mean fields) induced by the 
particle in this orbital \cite{AR.00}. Effective alignments are on average 
better reproduced in the covariant DFT (CDFT) calculations than in the cranked 
Nilsson-Strutinsky version of the MM approach based on phenomenological 
Nilsson potential, see comparisons presented in Refs.\ \cite{AR.00,ARR.99,AF.05}. 
Reasonable description of effective alignments can be obtained also in Skyrme 
DFT \cite{Eet.00,DD.95}, but it is somewhat plagued by the uncertainties 
in the definition of the coupling constants for time-odd mean fields 
\cite{DD.95,SDMMNSS.10}. This problem does not exists in the CDFT since 
time-odd mean fields are defined via Lorentz covariance \cite{VRAL.05}.

\begin{figure*}[ht]
\begin{center}
\includegraphics[width=14.0cm]{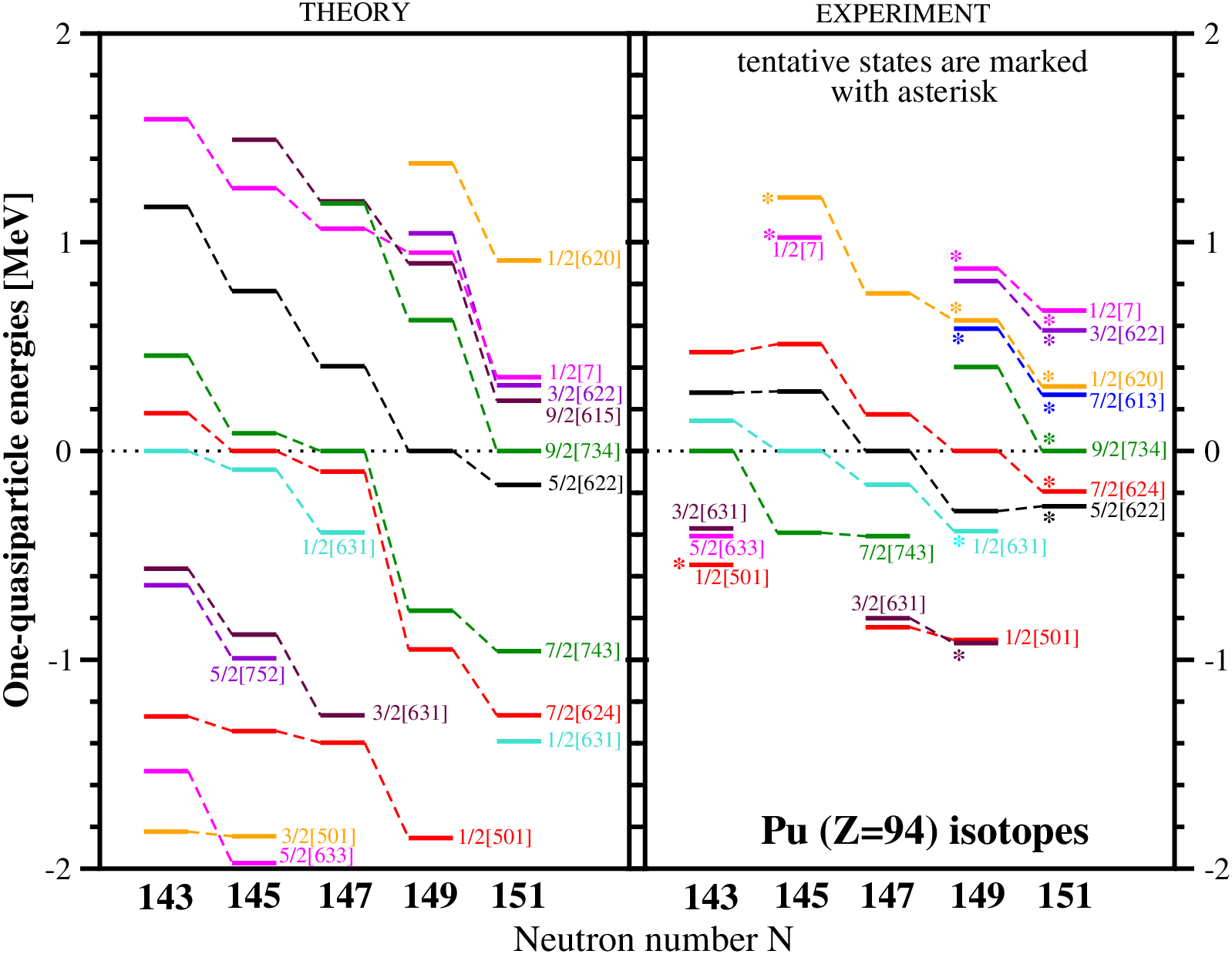}
\end{center}
\vspace{-0.5cm}
\caption{The evolution of one-quasineutron energies as a function of neutron 
number for the Pu isotopes. Hole states are plotted below the ground state 
(zero energy), and particle states are plotted above. Experimental data 
(one-quasineutron band-head energies) are taken from Ref.\ \cite{Eval-data}.
The squared amplitudes of the dominant component of the wave function change
gradually with neutron number. Thus, their variations are presented below 
in the neutron number range where these states are plotted in left panel
in the following format: [state label: A-B], where A is the squared amplitude
of the dominant component of wave function at lower value of $N$ and B at
upper value of $N$. These variations are $1/2[501]: 0.67-0.46$, $3/2[501]: 0.82-0.87$,
$5/2[752]: 0.57-0.56$, $5/2[633]: 0.77-0.78$, $3/2[631]: 0.55-0.57$, 
$7/2[743]: 0.72-0.74$, $1/2[632]: 0.55-0.57$, $7/2[743]: 0.72-0.74$, $1/2[631]: 0.58-0.51$,
$5/2[622]: 0.65-0.69$, $7/2[624]: 0.86-0.89$, $7/2[613]: 0.78-0.83$, 
$3/2[622]: 0.66-0.52$, $9/2[615]: 0.96-0.96$, $9/2[734]: 0.82-0.85$. The 1/2[7] 
state is strongly mixed. However, the cumulative squared amplitude of the 
components of the wave function with $N=7$ in the structure of this state 
exceeds 90\%. Thus, we label it only by principal quantum number $N$
and $\Omega$.}
\label{Eqp-evol-neu}
\end{figure*}

 The discussion above clearly shows that some aspects of the single-particle
motion are described on average better and in a more consistent way in 
self-consistent DFT than in phenomenological MM method. This is despite the 
fact that no single-particle information (apart of some spin-orbit splittings
in the case of Skyrme DFT) has been used in the fit of the DFT parameters, 
while the parameters of the Nilsson or Woods-Saxon potentials are fitted 
to experimental single-particle energies (see, for example, Refs.\ 
\cite{Beng85,GBI.86,DW.78,DMSWCN.79,CDNSW.87}). However, starting from earlier 
studies of the single-particle spectra of few nuclei in actinides within covariant 
\cite{A250} and Skyrme DFT \cite{BBDH.03}, and following by later global survey 
of the ground state configurations in odd-mass nuclei (Ref.\ \cite{BQM.07}) and 
the investigations of the spectra 
of odd-proton Ho nuclei (Fig. 6 in Ref.\ \cite{SDMMNSS.10}) in triaxial Skyrme 
DFT and the spectra of selected Rb \cite{RSR.10}, Y and Nb \cite{RSR.11} nuclei 
in axial Gogny DFT \cite{RSR.11}, it became clear that the single-particle 
spectra are poorly described with modern DFT. It is obvious that these theories
do not possess spectroscopic quality description of the single-particle spectra
which is achievable in the MM method as a consequence of the fit of the potential
parameters to experimental single-particle energies.

  Despite these results, no systematic analysis of the accuracy of the 
description of the single-particle spectra has been performed in DFT. The 
current manuscript aims on such analysis within the CDFT using extensive 
set of experimental data on the energies of one-quasiparticle states 
in deformed nuclei. There are two main goals behind of this study. 
First, typical uncertainties in the description of single-particle energies will 
be defined.  Second, the sources of discrepancies and possible ways of 
the improvement  of the description of the single-particle spectra will 
be discussed.

\section{The details of calculations}

   Cranked relativistic Hartree-Bogoliubov (CRHB) approach is used 
in the current manuscript in a similar way as it was done in an earlier 
study of the spectra of few odd-mass actinide nuclei in Ref.\ \cite{A250}.
Time-odd mean fields are taken into account, and the blocking due to
odd particle (see Sec.\ VI.A in Ref.\ \cite{A250} for details) is performed 
in a fully self-consistent way. The D1S Gogny force is used in the
pairing channel. Nuclear configurations of deformed odd nuclei (further 
one-quasiparticle [1-qp] configurations)  are labelled by means of the 
asymptotic quantum number $\Omega[N n_z \Lambda]$ (Nilsson  quantum number) of the 
dominant component of the wave function of blocked single-particle orbital. 
In each nucleus under study, the binding energies are calculated for a number 
of the 1-qp configurations based on the orbitals active in the vicinity of the 
Fermi level, and then the 1-qp spectra are built as shown in 
Fig.\ \ref{Eqp-evol-neu}.


  The CRHB equations are solved in the basis of an anisotropic 
three-dimensional harmonic oscillator in Cartesian coordinates. 
The same basis deformation $\beta_0=0.3$, $\gamma=0^{\circ}$ and 
oscillator frequency $\hbar \omega_0=41$A$^{-1/3}$ MeV have been 
used for all nuclei. All fermionic and bosonic states belonging 
to the shells up to $N_F=14$ and $N_B=18$ are taken into account 
in the diagonalization of the Dirac equation and the matrix 
inversion of the Klein-Gordon equations, respectively. 
The comparison with the results of calculations in a larger fermionic 
basis ($N_F=16$) shows that the energies of quasiparticle states are 
described with an accuracy of approximately $50$ keV which is sufficient 
for a statistical analysis.

\begin{figure}
\includegraphics[width=8.0cm]{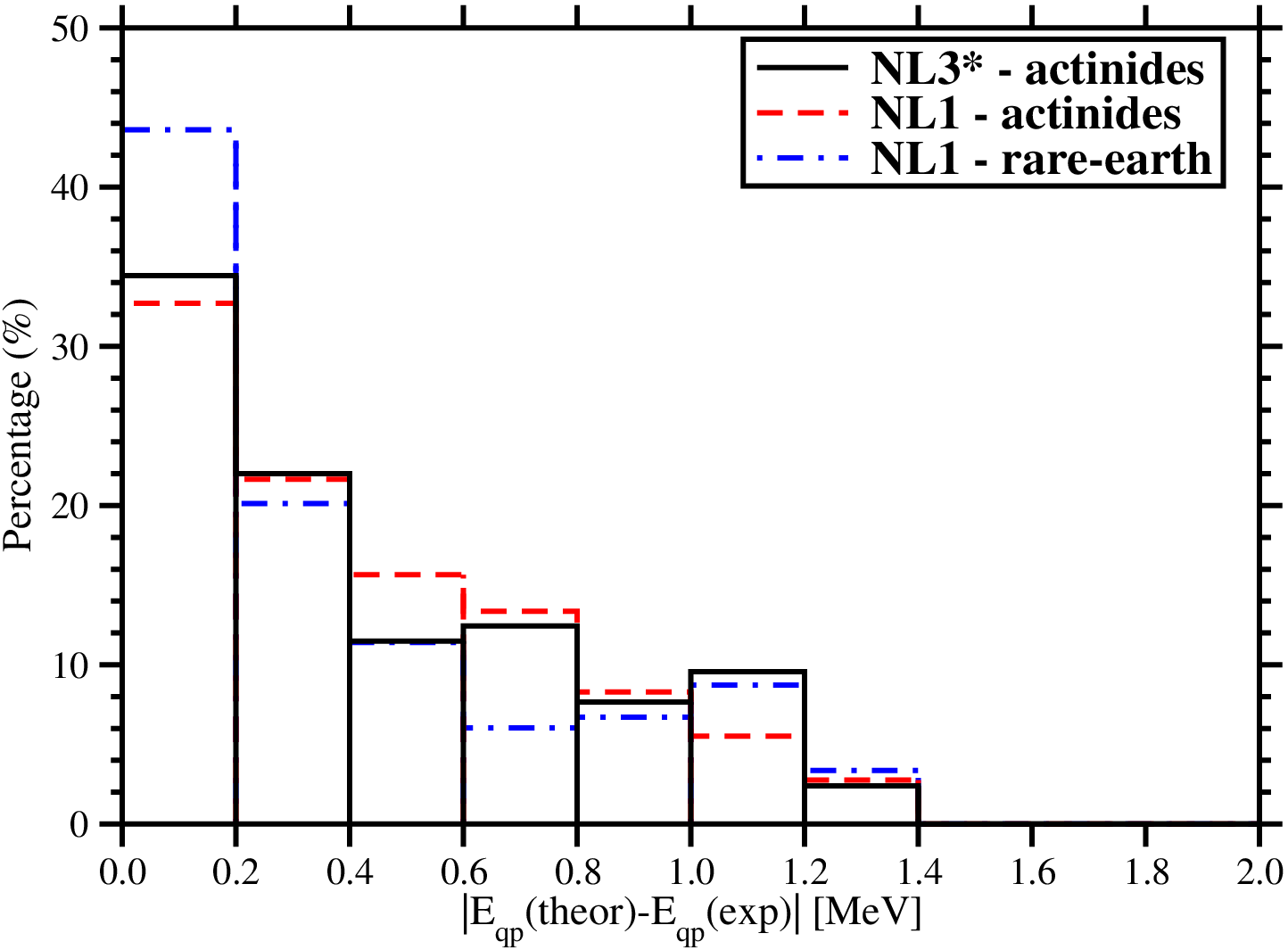}
\caption{The distribution of the deviations of the calculated energies 
$E_{qp}(theor)$ of one-quasiparticle states from experimental ones 
$E_{qp}(exp)$. The vertical axis shows the percentage of the states 
which deviate from experiment by the energy
deviation range (the width of bar) specified on horizontal axis. }
\label{Dev-stat}
\end{figure}

  The calculations have been performed with the NL1 \cite{NL1} and NL3* 
\cite{NL3*} parametrizations of the relativistic mean field (RMF) Lagrangian. 
The NL1 parametrization has been fitted to the nuclei in the valley of 
beta-stability. On the contrary, the fit of the NL3* parametrization includes 
neutron-rich nuclei so it is partially tailored towards the description of 
such nuclei. This recently fitted parametrization has been successfully 
applied to the description of binding energies \cite{NL3*}, ground state 
properties of deformed nuclei \cite{SRH.10}, spectra of odd spherical 
nuclei within the relativistic particle-vibration model \cite{LA.11},
rotating nuclei \cite{NL3*}, giant resonances \cite{NL3*}, and  breathing 
mode \cite{GLLM.10}. Note that only bulk properties of nuclei such as binding 
energies, radii etc. and nuclear matter properties have been used in the 
fit of these two parametrizations.

\begin{table}[h]
\caption{The summary of calculations. The number of calculated 
1-qp configurations and the number of experimental 1-qp states 
compared with calculations are displayed in second and third 
columns, respectively. Fourth column gives the percentage
of ground states, the structure of which is correctly reproduced
in the calculations.}
\label{Sn100-exp}
\begin{center}
\begin{tabular}{|c|c|c|c|} \hline
Region             &  calculated   & compared     & correct ground    \\ 
(parametrization)  &   states (\#) & states (\#)  &  states (\%)\\ \hline
 Actinides (NL3*)  & 415           &  209         &   38 \% \\
 Actinides (NL1)   & 444           &  217         &   45 \% \\
 Rare-earth (NL1)  & 360           &  149         &   48 \% \\ \hline
\end{tabular}
\end{center}
\label{Table-dev}
\end{table}

   1-qp spectra were calculated for 44 nuclei, namely, 21 odd-$N$ and 23 
odd-$Z$ nuclei in actinide region with $Z=89-100$. The nuclei with 
octupole deformation have been excluded from consideration. In addition, 
the calculations have been performed in rare-earth region. However, the 
experimental data on 1-qp states in rare-earth region are significantly 
larger than in actinide region \cite{Eval-data}. As a consequence, the 
calculations were performed only for odd-proton $Z=61$ (Pm), 63 (Eu), 65 
(Tb), 67 (Ho) and 69 (Tm) isotope chains and only with the NL1 parametrization 
of the RMF Lagrangian; in total, for 31 odd-proton nuclei. All these nuclei 
are deformed with quadrupole deformation $\beta_2 \geq 0.2$ and labelling 
of their single-particle states by means of the Nilsson quantum numbers is 
commonly accepted \cite{SFbook.74,BMbook.75,CAFE.77,JSSJ.90,NRbook,Sol-book2}. 
Table 1 provides the summary of these calculations. The data on adopted 
experimental one-quasiparticle levels are taken from Ref.\ \cite{Eval-data}.

\begin{figure}
\includegraphics[width=8.0cm]{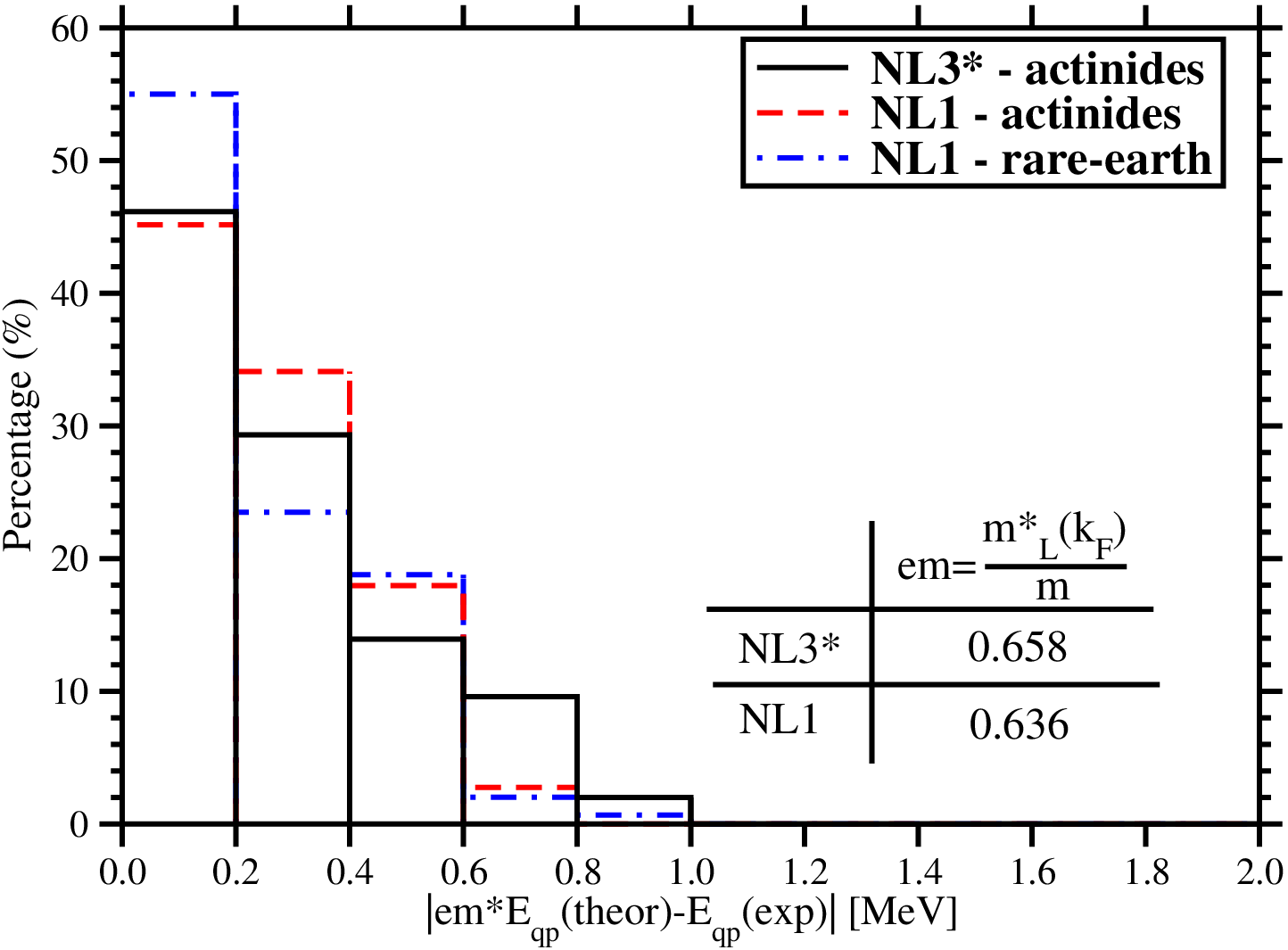}
\caption{The same as Fig.\ \ref{Dev-stat}, but for the
case when the energy scale of theoretical spectra is corrected for low 
Lorentz effective mass.}
\label{Dev-scaled}
\end{figure}

\section{Statistical analysis}

  Fig.\ \ref{Eqp-evol-neu} shows the comparison of calculated and experimental 
one-quasineutron  spectra in Pu isotopes. A number of features are clearly
seen.  First, for a given 1-qp state the discrepancy between theory and experiment
depends on neutron number. Second, for a given 1-qp state the slope of the energy 
curve as a function of neutron number is more pronounced in the calculations than 
in experiment. These two features are interconnected and they emerge from the fact 
that theoretical energy scale is more stretched out than experimental one 
due to low effective mass (see discussion below). The change of the Fermi 
energy with neutron number leads to the changes of the energy differences between 
ground and excited states and these differences are affected by effective mass in the 
calculations. Third, the relative energies of different experimental 
1-qp states are not always reproduced in calculations. This feature originates 
from the fact that the energies of spherical subshells, from which deformed states 
emerge, can deviate from experiment \cite{A250}). These three 
features are seen in all isotone and isotope chains.

     Note however that for a given state the 1-qp energy changes as a function 
of particle number are appreciably smaller when 1-qp energies of proton 
(neutron) subsystem are shown as a function of neutron (proton) number 
(see, for example, Fig.\ 6 in Ref.\ \cite{SDMMNSS.10}) because the changes 
in proton (neutron) Fermi energy and deformation are relatively small.

  Statistical analysis of the discrepancies between calculated and
experimental energies of one-quasiparticle states is presented in
Fig.\ \ref{Dev-stat}. One can see that in the actinide region only
approximately 33\% of one-quasiparticle states are described with an
accuracy better than 200 keV, and approximately 22\% with an accuracy   
between 200 and 400 keV. The percentage of states for a given range 
of deviations goes gradually down with an increase of the deviation 
between experiment and calculations. However, for some states the 
deviation of calculated energy from experimental one exceeds 1 MeV 
and can be close to 1.4 MeV. Note that earlier analysis  of the 
single-particle spectra of few actinide nuclei in Ref.\ \cite{A250} 
shows similar features.  Fig.\ \ref{Dev-stat} also shows that with the NL1 
parametrization the 1-qp energies in odd-proton rare-earth nuclei 
are somewhat better described as compared with actinide region. For 
example, the energies of 44\% of the states are described with an 
accuracy better than 200 keV. Otherwise, the distrubion histograms 
for the deviations are similar in both regions and for both 
parametrizations.

  Fourth column of Table \ref{Table-dev} shows the percentage of the 
ground states, the structure of which is correctly reproduced in the 
calculations. These values are comparable or somewhat better than the 
ones obtained in systematic Hartree-Fock+BCS calculations of deformed 
nuclei with SIII, SkM* and SLy5 Skyrme forces and FRDM calculations 
employing phenomenological folded-Yukawa potential \cite{BQM.07}, which 
show the agreement with experiment at about 40\% level.

  It is interesting that the overall accuracy of the description of 
the energies of deformed one-quasiparticle states is slightly better
in old NL1 parametrization, which was fitted 25 years ago mostly to 
the nuclei at the $\beta-$stability line, than in the recent NL3*
parametrization. This suggests that the inclusion of extra information 
on neutron rich nuclei into the fit of the NL3* parametrization may 
lead to some degradation of the description of single-particle states 
along the valley of beta-stability.

\section{The sources of the discrepancies between theory and experiment}

\subsection{Effective mass}
\label{Eff-mass}

  Low effective mass of CDFT is one of the sources of the discrepancies 
between theory and experiment. This is because the average level 
density of the single-particle states on the mean field level is 
related to the effective mass $m^*_L(k_F)/m$ (Lorentz mass in the 
notation of Ref.\ \cite{JM.89} for the case of CDFT) of the nucleons 
at the Fermi surface  which depends on momentum as \cite{JM.89}
\begin{eqnarray}
\frac{m^*_L(k_F)}{m} &=& \sqrt{\left(\frac{m^*(0)}{m}\right)^2 + 
\left(\frac{k_F}{m}\right)^2} \nonumber \\ 
& \approx & \sqrt{\left(\frac{m^*(0)}{m}\right)^2 + 0.08}
\end{eqnarray} 
where $m^*(0)/m$ is the value at $k=0$ which is called Dirac 
effective mass \cite{JM.89} and 0.08 is obtained for $(k_F/m)^2$
when typical 
value $k_F\approx 1.35$/fm is used. The values of Lorentz 
effective mass for employed parametrizations are given in 
Fig.\ \ref{Dev-scaled}.

  Low effective mass leads to a stretching of theoretical 
single-particle energy scale as compared with experiment (see 
Refs.\ \cite{LR.06,LA.11} for comparisons of calculated and 
experimental spectra in spherical nuclei). The role of the 
energy stretching due to low effective mass can
be illustrated on the example of the $\pi 1/2[420]$, $\pi 3/2[411]$
and $\pi 5/2[402]$ states in rare-earth region.  These states emerge 
from the $\pi d_{5/2}$ spherical subshell. For the majority of the 
nuclei under study, the $\pi 3/2[411]$ state is located close to the 
proton Fermi level. As a consequence, it is observed in 25 nuclei, 
and its energy in those nuclei is described in the calculations 
with an average accuracy of 250 keV. On the contrary, the $\pi 1/2[420]$ 
($\pi 5/2[402]$) state is located significantly below (above) the Fermi 
level both in experiment and calculations. However, stretching of 
theoretical energy scale due to low effective mass results in systematic
deviations of calculated energies of these states from experiment by 
more than 1 MeV. This leads to a peak in distribution histogram of the 
deviations of calculated energies from experimental ones at the 
deviation energy of around 1.1 MeV (see Fig.\ \ref{Dev-stat}). Another 
contributor to this peak is the $\pi 9/2[514]$ state.

  This stretching is also clearly visible when single-particle Nilsson 
diagrams obtained in the CDFT and phenomenological Nilsson
or Woods-Saxon potentials are compared (see, for example, Fig.\ 1 
in Ref.\ \cite{ARR.99}). These potentials are characterized by an 
effective mass $m^*(k_F)/m\approx 1.0$ which gives calculated 
level density close to experiment. As illustrated in spherical
nuclei, the calculated level density and single-particle spectra
are compressed and come closer to experimental ones when CDFT 
is suplemented by particle-vibrational coupling (PVC) 
\cite{LR.06,LA.11} (similar effect exists also in the PVC models 
based on non-relativistic DFT \cite{QF.78,MBBD.85,CSFB.10}).

  Similar compression of calculated spectra is expected also 
in deformed nuclei. However, no PVC model based on DFT framework 
has been developed so far for such nuclei.  Systematic comparison 
of single-particle Nilsson diagrams obtained in the CDFT and 
phenomenological Nilsson potential suggest that on average expected 
compression of the single-particle spectra can be achieved by 
multiplying the energies of the 1-qp configurations by Lorentz 
effective mass.  Such energy rescaling follows also from the
results of phenomenological scheme of Ref.\ \cite{VNR.02} based on 
a linear ansatz for the energy dependence of the scalar and vector 
components of the nucleon self-energy for the states close to the 
Fermi surface which simulates the dynamical effects that arise from 
the coupling of single-nucleon motion to collective surface 
vibrations.

  The impact of such energy rescaling on the distribution of 
the deviations between theory and experiment is shown in Fig.\ 
\ref{Dev-scaled}. One can see that more than 75\% of states are 
described with an accuracy better than 400 keV; this is a typical 
accuracy of the description of the energies of the deformed 1-qp 
states within phenomenological potentials \cite{JSSJ.90,PS.04}.
Although this energy rescaling is somewhat schematic, it clearly
indicates that PVC, leading to an increase of effective mass 
\cite{LR.06}, will also improve the description of experimental 
spectra as compared with mean field results; this has already been 
seen in spherical nuclei \cite{LA.11}.

\subsection{Relative energies of different states}

  Incorrectly calculated relative energies of different states
(see Fig.\ \ref{Eqp-evol-neu}) 
represent another source of the discrepancies between theory and 
experiment (see discussion of the Pu isotope chain above). It 
originates from the fact that relative energies of different 
spherical subshells, from which deformed states emerge, are not 
properly reproduced in model calculations \cite{A250}. This 
source shows up also in a statistical analysis. For example, the 
energy of the hole $\nu 1/2[501]$ state in actinide region is 
systematically higher than experimental one by around 1 MeV in 
both employed parametrizations. This is only deformed state 
originating from the $\nu p_{1/2}$ spherical subshell, so the 
current analysis suggests that in the calculations it is placed 
too deep (by approximately 1 MeV) with respect of other spherical 
subshells.

\begin{figure}[ht]
\includegraphics[width=8.0cm]{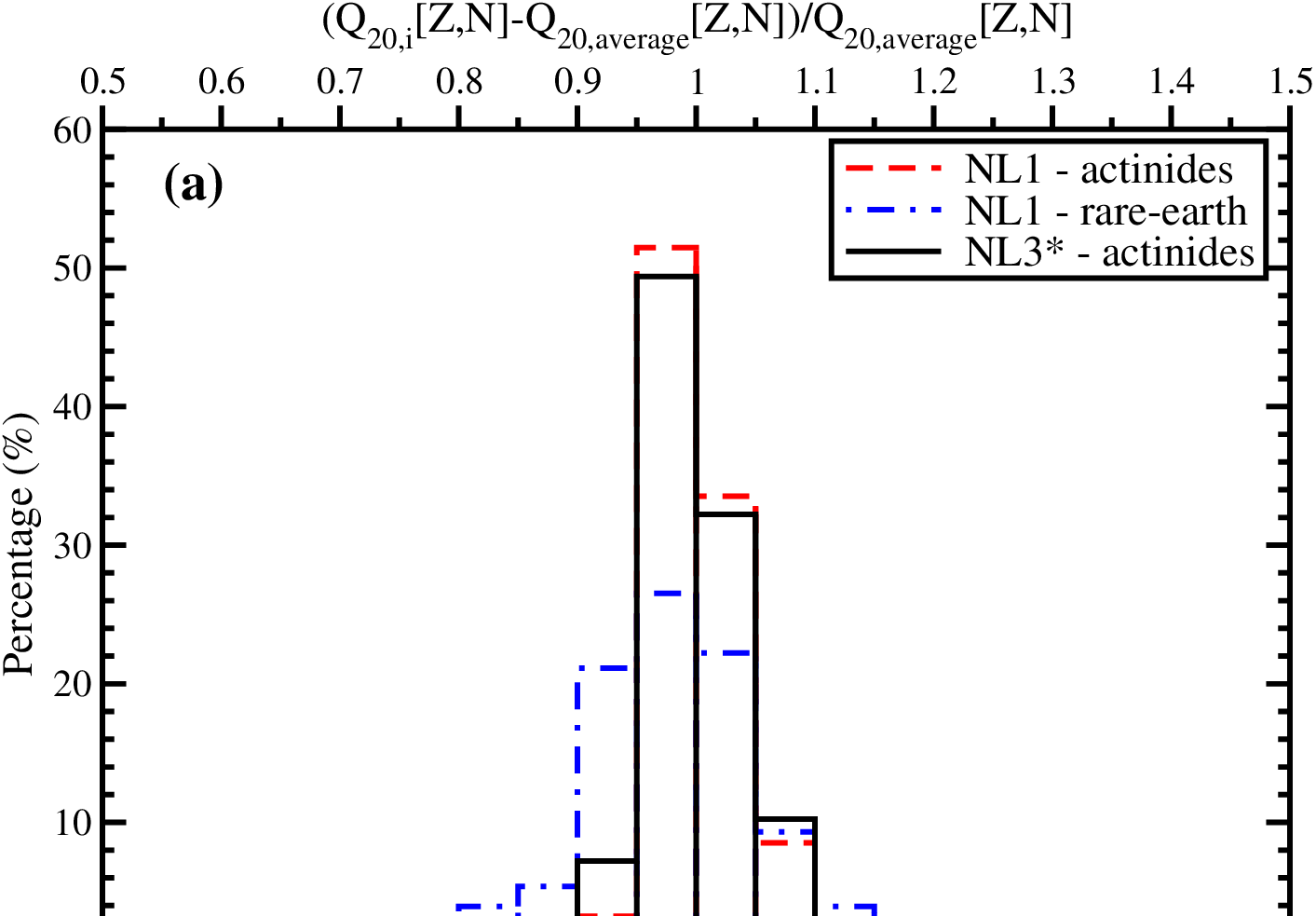}
\includegraphics[width=8.0cm]{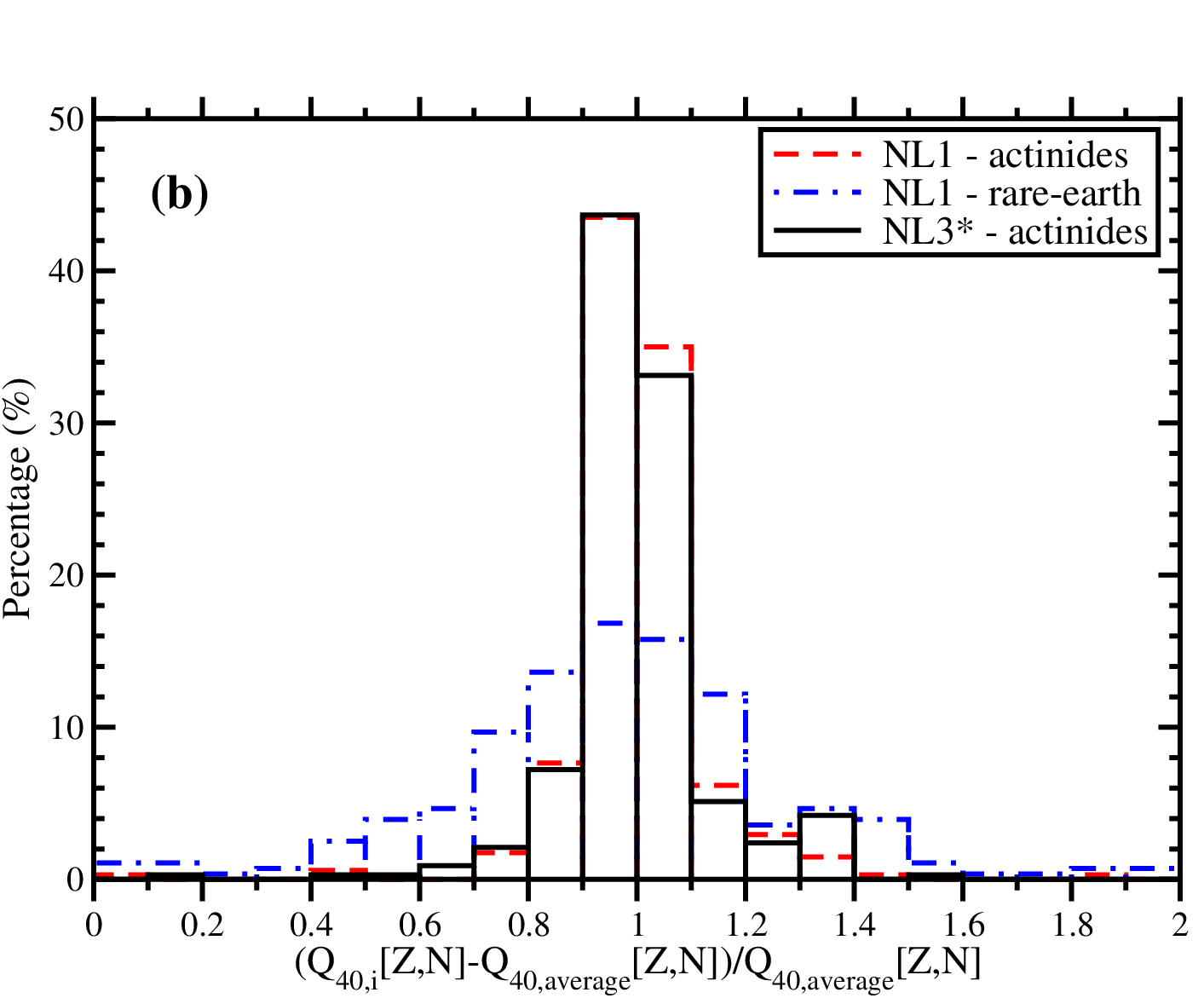}
\caption{Histogram of differences between the moment 
of the $i$-th 1-qp configuration and the average moment of all 
calculated 1-qp configurations in the $(N,Z)$ nucleus. The 
distributions for mass quadrupole (panel (a)) and mass hexadecapole
(panel (b)) moments are shown.}
\label{Beta-dev-dist}
\end{figure}

\section{Consequences for spectroscopic quality energy density functionals}

  The need for quantitative understanding and highly accurate description
of nuclear structure phenomena is a driving force behind the efforts to
develop spectroscopic quality energy density functionals (EDF). Such efforts 
are especially visible in the field of non-relativistic Skyrme DFT 
\cite{SDMMNSS.10,ZDSW.08}. However, the basic question whether this is 
possible on the mean-field level is still under debate \cite{BCS.10}.

  It is well known from Skyrme DFT that EDF, forced to describe accurately
single-particle spectra on the mean-field level, are characterized by 
effective mass close to one \cite{B.98}. In the CDFT, one cannot improve the 
description of the single-particle spectra by increasing effective mass since 
all CDFT parametrizations on the Hartree level have Lorentz effective mass 
$m^*_L(k_F)/m$ close to 0.65 \cite{VRAL.05}. The current analysis for deformed 
nuclei strongly suggests that further progress in improving spectroscopic 
quality of covariant EDF will be quite limited on the mean-field level. We
expect that even for other modern CDFT parametrizations, not employed in the 
current manuscript, the distributions of the deviations of calculated 
energies from experimental ones of the 1-qp states will be comparable or 
only slighthtly better than the one seen in Fig.\ \ref{Dev-stat}. This is 
due to their low Lorentz effective masses and unavoidable errors in relative 
placement of specific  single-particle orbitals. However, for some parametrizations such 
as NLSH and NL-RA1 obtained distributions can be even worse than that of
Fig.\ \ref{Dev-stat}; this follows from the analysis of Ref.\ \cite{A250}.

  As a consequence, the only way to substantially improve the description 
of the single-particle properties in the framework based on CDFT is to take 
into account PVC. It was already illustrated in spherical nuclei that this 
leads to a significant improvement in the description of the experimental 
energies of the dominant single-particle states \cite{LR.06,LA.11}.
In addition, it takes care of well-known fragmentation of the 
single-particle strength of the levels; this feature is completely ignored 
on the mean-field level. Based on experience in spherical nuclei, one can
expect that the inclusion of PVC in deformed nuclei will bring Lorentz 
effective mass close to one, thus leading to a level density which is similar to 
experimental one. This will definitely improve the description of the
1-qp spectra. The corrections to the energies of single-particle states due 
to PVC are strongly state-dependent in spherical nuclei \cite{LR.06,LA.11}. 
On the contrary, in deformed odd nuclei the corrections to the energies of 
1-qp states due to PVC are expected to be less state-dependent since surface
vibrations are more fragmented in deformed nuclei than in spherical ones 
\cite{Sol-book2}\footnote{This provides extra justification for the energy 
scaling procedure employed in Sect.\ \ref{Eff-mass}.}. However, they can still 
affect the relative energies of different deformed 1-qp states.

\section{Deformation effects}

  Each blocked single-particle orbital induces deformation polarization 
effects. Fig.\ \ref{Dev-scaled} shows the distribution of the mass 
quadrupole and hexadecapole moments of the 1-qp configurations relative
to the respective average moments of the full set of 1-qp configurations
in a given nucleus. This figure clearly shows that actinide nuclei are
reasonably rigid since polarization effects (especially for quadrupole
moments) induced by odd particle do not lead to substantial deviations
of equilibrium moments from average values. On the contrary, rare-earth 
nuclei are considerably softer than actinides since their histograms 
are significantly wider.  The results for actinides
show also that polarization effects do almost not depend on the parametrization
of the RMF Lagrangian; small differences between histograms obtained in
the NL1 and NL3* parametrizations are most likely due to slightly different 
sets of the 1-qp configurations obtained in the calculations. 
These changes in the moments/deformations and relevant changes in the binding
energies of the 1-qp configurations induced by odd particle have to be taken 
into account when experimental data are compared with the results of 
the calculations. However, they are completely ignored in the models most
frequently used for the analysis of the spectra of deformed odd nuclei such
as MM model \cite{PS.04},  particle+rotor \cite{NRbook} and quasiparticle-phonon
\cite{Sol-book2} models; these models  assume the same deformation for all 1-qp 
states in the nucleus under study.

  The calculations suggest that the deformation driving effects induced by 
odd proton or neutron are sufficient to polarize the nucleus towards 
positive or negative $\gamma$-deformation in some one-quasiparticle 
configurations. On average, only 6.8\% of calculated states have 
$\gamma$-deformation in the range $1^{\circ} \leq |\gamma| \leq 17^{\circ}$; 
the $\gamma$-deformation of other states is either exactly zero (in the majority 
of the cases) or below $1^{\circ}$. It turns out that the appearance of 
sizable $\gamma$-deformation correlates with blocking of few specific 
single-particle states. For example, in the actinide region 43\% of triaxial 
configurations has the dominant $\nu [622]3/2$ structure, and 38\% has 
the $\nu [501]1/2$ structure. Time-odd mean fields have almost no effect 
on the equilibrium deformations similar to previous CDFT studies. On the 
contrary, the results of Skyrme EDF  studies of Ref.\ \cite{SDMMNSS.10} 
show that the inclusion of time-odd mean fields favors the axial 
deformation of calculated configurations.

\section{Conclusions}

  In conclusion, for the first time a systematic analysis of the accuracy  
of the description of the energies of one-quasiparticle configurations 
in deformed odd nuclei has been performed in the DFT framework. It provides
theoretical estimates on the errors in calculated energies of one-quasiparticle 
configurations. Two sources of inaccuracies, namely, low effective mass 
leading to a stretching of the energy scale and incorrect relative 
positions of some single-particle states exist in model calculations. While 
the solution of the latter problem can be attempted in the DFT framework, 
we do not believe that it will lead to significant improvements in 
spectroscopic quality of energy density functionals. The comprehensive solution 
requires taking into account particle-vibration coupling which will (i) compress 
the calculated one-quasiparticle spectra bringing them closer to experiment and 
(ii) correct the relative energies of different single-particle states. In our 
opinion, only such approach combined with respective re-parametrization of the 
RMF Lagrangian can lead to spectroscopic quality energy density functionals.  

  In addition, our results show that one should be extremely careful in the
interpretation (predictions) of (for) the data which involves absolute or 
relative energies of different one-/two-/many-quasiparticle states since 
their energy description is associated with non-negligible theoretical 
errors. Extra care has to be taken also in the case when physical 
phenomenon sensitively depends on the single-particle energies. For 
example, the existence of very shallow left and right chiral minima with 
depth of around 200 keV separated by a tiny barrier ($\approx 60$ keV) is 
responsible for chiral rotation \cite{DFD.00}. Theoretical errors in the 
description of single-particle energies can either create false minima 
(barrier) or kill real ones.

\bigskip

{\leftline{\bf Acknowledgements} This work has been supported by the U.S. Department 
of Energy under the grant DE-FG02-07ER41459. }

\end{document}